\documentclass[12pt, a4paper] {article}

\usepackage{amsfonts}
\usepackage{textcomp}
\usepackage{color}
\usepackage{amsmath}
\usepackage{graphicx}
\usepackage{bm}
\usepackage{amssymb}
\usepackage{latexsym}
\usepackage{mathrsfs}
\usepackage{textcomp}
\begin{document}

\begin{center}

{\sf {\bf \large Anti Self-Dual Yang-Mills, Modified Faddeev-Jackiw Formalism and Hidden BRS Invariance }}

\vskip 1.0 cm

{\sf{ \bf Saurabh Gupta and Raju Roychowdhury}}\\
\vskip .1cm
{\it Instituto de F\'\i sica, Universidade de S\~ao Paulo,\\
C. Postal 66318,  05314-970 S\~ao Paulo, SP, Brazil}\\

\vskip .15cm
{E-mails: {\tt saurabh@if.usp.br; raju@if.usp.br}}

\end{center}

\vskip 1 cm

\noindent
{\bf Abstract:} We analyze the constraints for a system of anti self-dual Yang-Mills (ASDYM)
equations by means of the modified Faddeev-Jackiw method in $K$ and $J$ gauges {\it \`{a} la} Yang. 
We also establish the Hamiltonian flow for ASDYM system through the hidden BRS invariance in both 
the gauges. Finally, we remark on the bi-Hamiltonian nature of ASDYM and the compatibility of the symplectic structures therein. 
\\



\vskip 1.0 cm
\newpage

\section{Introduction}

From the point of view of integrable systems, a set of equations that is of particular interest is the anti self-dual Yang-Mills (ASDYM) equations 
which is a system of partial differential equations defined on an arbitrary oriented four manifold $\mathbb{CM}$ and depends on the choice of a 
Lie group. If the four manifold turns out to be anti self-dual then the ASDYM equations are integrable in general since there exists a twistor 
theoretic method for constructing solutions that is comparable to the inverse scattering methods widely known in the literature of integrable 
systems \cite{AHS, MW, WW, Das}. It is well known that many of the integrable systems can be derived as symmetry reductions of ASDYM equations 
on an anti self-dual manifold for particular choices of the gauge group \cite{MW,paul}.
 
In double-null coordinates $(z, \tilde z, \omega, \tilde \omega)$  the metric on $\mathbb{CM}$ is defined as: 
\begin{eqnarray}
 ds^2 = 2 (dz d \tilde z - d\omega d \tilde \omega),
\end{eqnarray}
and the coordinates take the following form
\begin{eqnarray}
&& z = \frac {1}{\sqrt 2} (x^0 - i x^1), \quad  \omega = \frac {1}{\sqrt 2} (-x^2 + i x^3), \nonumber\\
&& \tilde z = \frac {1}{\sqrt 2} (x^0 + i x^1), \quad  \tilde \omega = \frac {1}{\sqrt 2} (x^2 + i x^3),
\end{eqnarray}
where $(x^0, x^1, x^2, x^3)$ are real Cartesian coordinates. Thus, in double null coordinates, the anti self-dual condition on $F = F_{ab} dx^a \wedge dx^b$  \footnote{Remember that $F_{ab} = \partial_a \Phi_b - \partial_b \Phi_a +[\Phi_a , \Phi_b]$} can be expressed as 
\begin{eqnarray}
 [D_z, D_\omega] = 0, \quad [D_{\tilde z}, D_{\tilde\omega}] = 0, \quad [D_z, D_{\tilde z}] - [D_\omega, D_{\tilde\omega}] = 0,
\end{eqnarray}
where $D = d + \Phi$ is defined to be the connection on a complex rank-n vector bundle $E$ in some local coordinate patch $U$ and $F$ is its curvature 2-form.
Also remember that the subscripts denote the partial derivative and so the components of the connections are
\begin{eqnarray}
 D_\omega = \partial_\omega + \Phi_\omega, \quad D_z = \partial_z + \Phi_z, \quad D_{\tilde\omega} = \partial_{\tilde\omega} + \Phi_{\tilde\omega}, 
 \quad D_{\tilde z} = \partial_{\tilde z} + \Phi_{\tilde z}.
\end{eqnarray}
ASDYM being an integrable system, the Lax pair can be constructed as $L = D_\omega - \xi D_{\tilde z}, \; \;  M = D_z - \xi D_{\tilde \omega}$, which commutes with each other for any arbitrary $\xi$.

It is well known that ASDYM equations can be branded into a Lagrangian formalism \`{a} la Yang \cite{yang,nair}.  
An explicit form of the symplectic structure for ASDYM equations in Yang's $K$ and $J$ gauges has been derived and the 
bi-Hamiltonian structure has also been established \cite{nut}. In the paper, the authors have also pointed out the existence of the constraints in both the gauges \`{a} la Dirac.
Dirac formalism \cite{dirac} - the standard method of quantization has been widely used in order to quantize Hamiltonian systems having 
constraints. In the Dirac approach, constraints are classified as primary, secondary, tertiary etc., or first-class and second-class. 
These constraints can be weak or strong (see, e.g. \cite{hen} for details).

However, there is yet another approach by Faddeev and Jackiw \cite{FJ} which is geometrically motivated and based on the symplectic structures 
for the quantization of singular systems. It deals with the Lagrangians that are first-order in nature. The classification of a constrained 
system is related to the singular behavior of the symplectic two-form and the brackets are obtained directly from the elements of the inverse 
of the symplectic two-form matrix \cite{BN,BN1}. However, the two approaches \`{a} la Dirac and Faddeeev-Jackiw are consistent with each other \cite{garcia}.  

The purpose of this article is twofold. Firstly to review the relevant literature and results on various aspects of ASDYM equations and then 
do the full constarint analysis for ASDYM in both $K$ and $J$ gauges \`{a} la Yang \cite{yang} using the modified Faddeev-Jackiw approach and derive 
the corresponding symplectic matrix and find all the constraints of the system at one go. The second interesting aim of our paper is to derive 
the Hamiltonian flow for ASDYM system through the hidden BRS invariance \`{a} la Gozzi et.al.\cite{hidden1, hidden2, GRT}.

The plan of the paper is as follows. In Section 2, we give a brief synopsis of the modified Faddeev-Jackiw formalism to quantize a system with 
constraints. Section 3 is devoted to the derivation of the full set of constraints for a system of ASDYM equations, within the framework of 
modified Faddeev-Jackiw formalism, in both  $K$ and $J$ gauges. We establish Hamiltonian flow through hidden BRS invariance and remark about 
the bi-Hamiltonian structure and compatibility of symplectic structures in Section 4. In Section 5, we make concluding remarks and point out some future directions.
Finally, the Appendix deals with the explicit calculation of the second-iterative symplectic matrix and derivation of new constraints in $J$ gauge.

\section{Modified Faddeev-Jackiw symplectic formalism}
In this section we give a brief synopsis of modified Faddeev-Jackiw approach (see, e.g. \cite{FJ,BN,BN2,mon,hua1,hua2}) to quantize a system with constraints. Let any n-dimensional manifold, in the 
configuration space, is described by the coordinates $\xi_i, i = 1, 2,..., n$. Then, one can construct a first-order symplectic Lagrangian density as 
\begin{eqnarray}
 {\cal L} = a_i (\xi) \; \dot \xi^i - V^{(0)} (\xi), \label{fjl}
\end{eqnarray}
where $V^{(0)}(\xi)$ is the symplectic potential and $a_i (\xi)$ are arbitrary one-form components. The Euler-Lagrange equations of motion for (\ref{fjl})
can be written as 
\begin{eqnarray}
 f_{ij}^{(0)} \; \dot \xi^j = \frac{\partial V^{(0)} (\xi)}{\partial \xi^i}, \label{eqm}
\end{eqnarray}
with $f$ being the symplectic two-form defined as $f = d a \equiv \frac{1}{2} f_{ij}^{(0)} d \xi^i d \xi^j$  and the symplectic matrix $f_{ij}^{(0)}$ is given in 
the following manner
\begin{eqnarray}
 f_{ij}^{(0)} = \frac{\partial a_j}{\partial \xi^i} - \frac{\partial a_i}{\partial \xi^j}.
\end{eqnarray}
If, the inverse of $f_{ij}^{(0)}$ exists, i.e. the symplectic matrix is nonsingular. Then solutions of (\ref{eqm}) can be obtained, as 
\begin{eqnarray}
 \dot \xi^j = (f_{ij}^{(0)})^{-1} \; \frac{\partial V^{(0)}(\xi)}{\partial \xi^i}.
\end{eqnarray}
However, if the symplectic matrix $f_{ij}^{(0)}$ turns out to be singular, it implies that the system is endowed with constraints. Then, 
according to the prescription described in \cite{BN,BN2}, we need to find the zero-modes of the singular matrix. Then the constraints $(\Omega^{(0)})$ 
can be obtained from 
\begin{eqnarray}
 \Omega^{(0)}_\alpha = (v_i^{(0)})^T_\alpha \; \frac{\partial V^{(0)} (\xi)}{\partial \xi^{i}} = 0,
\end{eqnarray}
where $(v_i^{(0)})$ are the zero-modes of the symplectic matrix. 
Until now, the procedure is similar to the usual Faddeev-Jackiw method \cite{FJ,BN,BN2}. Now, we describe modified Faddeev-Jackiw method to derive new 
constraints \cite{hua1,hua2}.
In the modified method, the consistency condition analogous to Dirac-Bergmann approach is used to derive new constraints, as prescribed below
\begin{eqnarray}
 \dot \Omega^{(0)} = \frac{\partial \Omega^{(0)}}{\partial \xi^i} \; \dot \xi^i = 0. \label{17}
\end{eqnarray}
Combining (\ref{17}) with  (\ref{eqm}), we have 
\begin{eqnarray}
 f_{kj}^{(1)} \; \dot \xi^j \; = \; Z_k (\xi),
\end{eqnarray}
where
\begin{eqnarray}
 f_{kj}^{(1)} = \left( \begin{array}{c}
 f_{ij}^{(0)} \\
 \frac{\partial \Omega^{(0)}}{\partial \xi^i}
\end{array}
\right) \qquad {\text and} \qquad 
Z_k(\xi) = \left(\begin{array}{c}
   \frac{\partial V^{(0)}(\xi)}{\partial \xi^i} \\
   0
\end{array}
\right).
\end{eqnarray}
The new constraints can be deduced in the following fashion
\begin{eqnarray}
 (v^{(1)})_k^T \; Z_k |_{\Omega^{(0)} = 0}  \; = \; 0,
\end{eqnarray}
where $v^{(1)}_k$ are the zero modes of the matrix $f_{kj}^{(1)}$. The above equation is then evaluated at $\Omega^{(0)} = 0$. If 
it turns out to be an identity (i.e. 0 = 0), then there are no further constraints otherwise it will lead to the  constraints 
$\Omega^{(1)}$  given by 
\begin{eqnarray}
 \Omega^{(1)} = (v^{(1)})_k^T \; Z_k |_{\Omega^{(0)} = 0}  \; = \; 0.
\end{eqnarray}
Similarly one can now introduce the consistency condition for $\Omega^{(1)}$, as
\begin{eqnarray}
  \dot \Omega^{(1)} = \frac{\partial \Omega^{(1)}}{\partial \xi^i} \; \dot \xi^i = 0, 
\end{eqnarray}
and combine it with (\ref{eqm}) and (\ref{17}) in order to deduce new constrains, if any. These steps have to be repeated till there are no 
further constraints.

\section{Constraint analysis of ASDYM: Modified Faddeev-Jackiw approach}

It is well established that the ASDYM equations can be recast into a Lagrangian formalism, one is due to Yang \cite{yang,nair} and another
expression by Leznov \cite{lez} and Parkes \cite{par}. In this section, we apply the modified Faddeev-Jackiw approach to derive the 
full constraints of ASDYM system in Yang's framework. For this purpose, we considered both the gauges, i.e. $K$-gauge and $J$-gauge {\it \`{a} la} Yang,
in our present endeavor.

\subsection{K-gauge}
In $K$-gauge, {\it \`{a} la} Yang \cite{yang}, the ASDYM equations can be derived from the following $K$-gauge Lagrangian \cite{nut,mukht}
\begin{eqnarray}
 {\cal L_K} = \frac{1}{2} K_z K_{\tilde z} - \frac {1}{2} K_\omega K_{\tilde \omega} + \frac {2}{3} K [K_\omega, K_z].
\end{eqnarray}
Here, we consider the independent variable $z$ as `time' and the trace operation is assumed. The first-order Lagrangian, in $K$-gauge, can be expressed as \cite{nut}
\begin{eqnarray}
 {\cal L}^{I}_{\cal K} = \frac{1}{2} M \tilde M + \frac {1}{2} K_\omega K_{\tilde \omega} - \frac {1}{2} \tilde M K_z - \frac {1}{2} M K_{\tilde z} 
 + \frac{1}{3} M [K, K_\omega], \label{folk}
\end{eqnarray}
where $M$ and $\tilde M$ are the newly introduced variables (double in number as they are required) due to the asymmetry present in the 
anti self-duality condition between the independent variables and their complex conjugates (see, e.g. \cite {nut} for details).
The Euler-Lagrange equations of motion, derived from the above Lagrangian, can be written as
\begin{eqnarray}
&& \tilde M = K_{\tilde z} + \frac{2}{3} [K_\omega, K], \qquad M = K_z, \nonumber\\
&& \tilde M_z = M_{\tilde z} + \frac {2}{3} \big([M_\omega, K] + [K_\omega, M]\big). \label{EOMK} 
\end{eqnarray}

We can identify $X^1 \equiv \tilde M, X^2 \equiv K$ and $X^3 \equiv M$ as the phase space coordinates.  
The corresponding canonically conjugate momenta can be defined in the following fashion 
\begin{eqnarray}
 \Pi _A \equiv \frac{\partial {\cal L}^I_{\cal K}}{\partial X^A_z},
\end{eqnarray}
where $A = 1,2,3$ and subscript $z$ represents `time' derivative. The canonically conjugate momenta are listed below:
\begin{eqnarray}
 \Pi_{\tilde M} \approx 0, \qquad \Pi_K = - \frac{1}{2} \tilde M, \qquad \Pi_{ M} \approx 0. \label{cons}
\end{eqnarray}
Before moving towards the modified Faddeev-Jackiw approach we would like to mention few noteworthy observations. First, it is straightforward to 
check that the first-order Lagrangian (\ref{folk}) is singular as its Hessian vanishes identically. 
Hence, it indicates the existence of the constraints in the language of Dirac \cite{dirac}. Second, the first two constraints (cf. (\ref{cons}))  
are second-class in nature as they have non-zero Poisson brackets among themselves, where as the last one is first-class in nature \cite{nut}.

To apply Faddeev-Jackiw approach we need to express the Lagrangian density in first-order symplectic form. For this purpose, we calculate following 
\begin{eqnarray}
 \Pi_A \; X^A_z - {\cal L}^{I}_{\cal K} \; = \;  \Pi_K \; K_z - \frac{1}{2} M \tilde M - \frac {1}{2} K_\omega K_{\tilde \omega} + \frac {1}{2} \tilde M K_z 
 + \frac {1}{2} M K_{\tilde z} - \frac{1}{3} M [K, K_\omega].
\end{eqnarray}
So, the first-order symplectic Lagrangian density, in $K$-gauge, is given by
\begin{eqnarray}
 {\cal L}^{(0)}_{\cal K} = \Pi_A \; X^A_z - V^{(0)}_{\cal K}, 
\end{eqnarray}
where $V^{(0)}_{\cal K}$ is the symplectic potential and given by 
\begin{eqnarray}
 V^{(0)}_{\cal K} = - \frac{1}{2} M \tilde M - \frac {1}{2} K_\omega K_{\tilde \omega} 
 + \frac {1}{2} M K_{\tilde z} - \frac{1}{3} M [K, K_\omega].
\end{eqnarray}
The corresponding symplectic equations of motion can be deduced from the following expression 
\begin{eqnarray}
 f_{AB}^{(0)} \; \xi^B_z \; = \; \frac{\partial V^{(0)}_{\cal K}}{\partial \xi^A},     \label{fab}
\end{eqnarray}
where 
\begin{eqnarray}
 f_{AB}^{(0)} \; (\omega, \tilde \omega) \; = \; \frac{\delta a_B (\tilde\omega)}{\delta \xi^A (\omega)} 
 - \frac{\delta a_A (\omega)}{\delta \xi^B (\tilde\omega)}. \label{fab1}
\end{eqnarray}
The symplectic variable set is given as follows 
\begin{eqnarray}
 \xi^{(0)}_{\cal K} (\omega) = \{\tilde M, K, \Pi_K, M \}.
\end{eqnarray}
With this we can calculate the components of symplectic 1-form as listed below:
\begin{eqnarray}
 a^{(0)}_{\tilde M} = 0, \quad a^{(0)}_K = -\frac{1}{2} {\tilde M} \equiv \Pi_K, \quad a^{(0)}_{\Pi_K} = 0, \quad a^{(0)}_M = 0.
\end{eqnarray}
Thus, we obtain following symplectic matrix 
\begin{eqnarray}
 f_{AB}^{(0)} \; (\omega, \tilde \omega) \;  = \;  \left(
                                            \begin{array}{cccc}
                                                     0 & 0 & 0 & 0 \\
                                                     0 & 0 & -1 & 0 \\
                                                     0 & 1 & 0 & 0 \\
                                                     0 & 0 & 0 & 0
                                                    \end{array}
                                                    \right)    \delta (\omega - \tilde \omega),
\end{eqnarray} 
which is a singular matrix. The zero-modes of this matrix are $(v^{(0)})_1^T = (v_{\tilde M}^{(0)},0,0,0)$ and  
$(v^{(0)})_2^T = (0,0,0,v_{M}^{(0)})$, where $v_{\tilde M}^{(0)}, v_{M}^{(0)}$ are arbitrary constants. In the
view of Faddeev-Jackiw method, these zero modes will lead to the following constraints, as
\begin{eqnarray}
 \Omega^{(0)} \; &=& \;  (v^{(0)})_A^T \; \frac{\partial V^{(0)}_{\cal K}}{\partial \xi^A}  \nonumber\\
                 &=& v_{\tilde M}^{(0)} \Big(\frac{-1}{2} M \Big) + v_{M}^{(0)} \Big(\frac{-1}{2} \tilde M 
                 + \frac{1}{2} K_{\tilde z} - \frac{1}{3} [K, K_\omega] \Big) \approx 0.
\end{eqnarray}

In order to derive new constraints, we shall take recourse to the modified Faddeev-Jackiw method. According to the 
prescription, as mentioned in the last section, we have
\begin{eqnarray}
 \Omega_z^{(0)} \; = \; \frac{\partial \Omega^{(0)}}{\partial \xi^A} \; \xi^A_z \; = \; 0. \label{ogdot}
\end{eqnarray}
Combining (\ref{ogdot}) with (\ref{fab}) and reformulating the combined equation, we have 
\begin{eqnarray}
 f_{CD}^{(1)} \; \xi^D_z \; = \; Z_C (\xi),           \label{fcd}
\end{eqnarray}
where
\begin{eqnarray}
 f_{CD}^{(1)} &=&  \left ( \begin{array}{c}
                          f_{AB}^{(0)} \\
                          \frac{\partial \Omega^{(0)}}{\partial \xi^A}
                         \end{array}
                 \right) 
              =     \left(
                     \begin{array}{cccc}
                     0 & 0 & 0 & 0 \\
                     0 & 0 & -1 & 0 \\
                     0 & 1 & 0 & 0 \\
                     0 & 0 & 0 & 0 \\
                     -\frac{1}{2} & 0 & 0 & -\frac{1}{2}
                    \end{array}
                    \right)    \delta (\omega - \tilde \omega), \nonumber\\
                    {\text and} \nonumber\\ 
Z_C (\xi) &=& \left(\begin{array}{c}
                   \frac{\partial V^{(0)}_{\cal K}}{\partial \xi^A} \\
                   0
                  \end{array}
             \right) 
          =  \left( 
             \begin{array}{c}
              \frac{-1}{2} M\\
              \frac{-1}{3} [K_\omega, M] \\
              0\\
              \frac{-1}{2}\tilde M + \frac{1}{2} K_{\tilde z} - \frac{1}{3} [K, K_\omega]\\
              0
             \end{array}
            \right). 
\end{eqnarray} 
A closer look at $f_{CD}^{(1)}$ reveals that it is not a square matrix but still has a zero mode $(v^{(1)})^T = (-v^{(1)}_{\tilde M},0,0,v^{(1)}_M)$,
where $v^{(1)}_{\tilde M}, v^{(1)}_M $ are arbitrary constants. 
Multiplying this zero mode to  both sides of (\ref{fcd}) and evaluating at $\Omega^{(0)} = 0$,  we get the new constraints (if any). Mathematically, 
it can be expressed as:
\begin{eqnarray}
        Z_C (\xi) \;  (v^{(1)})^T_C |_{\Omega^{(0)} = 0} \; = \; 0.   \label{omg1}
\end{eqnarray}
The LHS of the above relationship implies 
\begin{eqnarray}
         \left( \begin{array}{cccc}
                 \frac{M}{2} & 0 & 0 & \frac{-M}{2}\\
                 \frac{1}{3} [K_\omega, M] & 0 & 0 & - \frac{1}{3} [K_\omega, M]\\
                 0 & 0 & 0 & 0 \\
                 \frac{1}{2}\tilde M - \frac{1}{2} K_{\tilde z} + \frac{1}{3} [K, K_\omega] & 0 & 0 &  \frac{-1}{2}\tilde M + \frac{1}{2} K_{\tilde z} - \frac{1}{3} [K, K_\omega] \\
                 0 & 0 & 0 & 0
                \end{array}
         \right), 
\end{eqnarray}
which is identically zero on the constraint surface $\Omega^{(0)} = 0$ and (\ref{omg1}) are identities 0 = 0.  
Thus, there are no further constraints in the system.

\subsection{J-gauge}

The explicit expression for the second-order Lagrangian, in J-gauge and for $SU(2)$ gauge group, is given by Pohlmeyer \cite{pol} as 
\begin{eqnarray}
 {\cal L}_{J} = \frac{1}{2 \phi^2} \Big(\phi_z \phi_{\tilde z} - \phi_\omega \phi_{\tilde\omega} + \bar\rho_z \rho_{\tilde z}
 - \bar\rho_\omega \rho_{\tilde \omega} \Big).
\end{eqnarray}
Here Yang's parametrization of $J$ matrix in terms of Poincar\'{e} coordinates $\rho$ and $\phi$ (with $\phi$ real and $\rho$ complex, 
see e.g. \cite{nut} for details) has been used. The above Lagrangian density can also be recast into the following first-order form, as \cite{nut}
\begin{eqnarray}
 {\cal L}^I_J = -\frac{1}{2} P \bar P + \frac{1}{2\phi} (P \phi_{\tilde z} + \bar P \phi_z) - \frac{1}{2 \phi^2} (\phi_\omega \phi_{\tilde \omega} + 
 \rho_{\tilde \omega} \bar\rho_\omega) - \frac{1}{2 \phi^2} (Q \bar Q - Q \bar\rho_z -\bar Q \rho_{\tilde z}).
\end{eqnarray}
The Euler-Lagrange equations of motion arising from the above Lagrangian are as follows
\begin{eqnarray}
&& P = \phi^{-1} \phi_z, \qquad \bar P = \phi^{-1} \phi_{\tilde z}, \qquad Q = \rho_{\tilde z}, \qquad \bar Q = \bar \rho_z, \nonumber\\
&& (\phi^{-2} \rho_{\tilde \omega})_\omega - (\phi^{-2} Q)_z = 0, \quad (\phi^{-2} \bar\rho_{\omega})_{\tilde \omega} - (\phi^{-2} \bar Q)_{\tilde z} = 0, \nonumber\\
&& P_{\tilde z} + \bar P_z - 2 \phi^{-1} \phi_{\omega \tilde\omega} + 2 \phi^{-2} \phi_\omega \phi_{\tilde \omega} - 2 \phi^{-2} \rho_{\tilde \omega} \bar \rho_\omega
+ 2 \phi^{-2} Q \bar Q = 0.
\label{EOMJ}
\end{eqnarray}
The canonically conjugate momenta are defined in the following fashion
\begin{eqnarray}
 \Pi_\Phi = \frac{\partial {\cal L}^I_J}{\partial \Phi_z}; \qquad {\text where} \quad \Phi = \bar \rho, \bar P, Q, \phi, P, \bar Q, \rho,
\end{eqnarray}
which yields following 
\begin{eqnarray}
 \Pi_{\bar \rho} = \frac{Q}{2 \phi^2}, \quad \Pi_\phi = \frac{\bar P}{2 \phi}, \quad \Pi_{\bar P} = 0, \quad \Pi_Q = 0, \quad \Pi_P = 0, 
\quad \Pi_{\bar Q} = 0, \quad \Pi_\rho = 0. \label{consj}
\end{eqnarray}
It is straightforward to check, using Dirac's prescription for constraint analysis \cite{dirac}, that last three constraints listed above 
(cf. (\ref{consj})) are first-class in nature as they have vanishing Poisson brackets and the rest of the constraints are second-class \cite{nut}. 
Now, to proceed with Faddeev-Jackiw approach the $J$-gauge Lagrangian is expressed in the first-order symplectic form in the following manner 
\begin{eqnarray}
{\cal L}^{(0)}_J \; = \; \Pi_\Phi \Phi_z - V^{(0)}_J,
\end{eqnarray}
where
\begin{eqnarray}
V^{(0)}_J \; = \; \frac{1}{2} P \bar P - \frac{1}{2\phi} (P \phi_{\tilde z}) + \frac{1}{2 \phi^2} (\phi_\omega \phi_{\tilde \omega} + 
 \rho_{\tilde \omega} \bar\rho_\omega) + \frac{1}{2 \phi^2} (Q \bar Q - \bar Q \rho_{\tilde z}). 
\end{eqnarray}
The corresponding symplectic equations of motion can be deduced by using (\ref{fab}) and (\ref{fab1}) with the set of following symplectic
variables 
\begin{eqnarray}
 \xi^{(0)}_J \; = \; \{\bar \rho, \Pi_{\bar \rho}, \bar P, Q, \phi, \Pi_\phi, P, \bar Q, \rho \}, 
\end{eqnarray}
and the components of symplectic 1-form as
\begin{eqnarray}
&& a^{(0)}_{\bar \rho} = \frac{Q}{2 \phi^2} \equiv \Pi_{\bar \rho}, \qquad a^{(0)}_\phi = \frac{\bar P}{2 \phi} \equiv \Pi_\phi, \qquad 
 a^{(0)}_{\Pi_{\bar \rho}} = 0,  \qquad a^{(0)}_{\bar P} = 0,   \nonumber\\
&& a^{(0)}_Q = 0, \qquad a^{(0)}_{\Pi_\phi} = 0, \qquad a^{(0)}_P = 0, \qquad a^{(0)}_{\bar Q} = 0, \qquad a^{(0)}_\rho = 0. 
\end{eqnarray}
Thus, we obtain following symplectic matrix
\begin{eqnarray}
 f_{AB}^{(0)} \; (\omega, \tilde \omega) = \left( \begin{array}{ccccccccc}
                        0 & -1 & 0 & 0 & 0 & 0 & 0 & 0 & 0 \\
                        1 & 0 & 0 & 0 & 0 & 0 & 0 & 0 & 0 \\
                        0 & 0 & 0 & 0 & 0 & 0 & 0 & 0 & 0 \\
                        0 & 0 & 0 & 0 & 0 & 0 & 0 & 0 & 0 \\
                        0 & 0 & 0 & 0 & 0 & -1 & 0 & 0 & 0 \\
                        0 & 0 & 0 & 0 & 1 & 0 & 0 & 0 & 0 \\
                        0 & 0 & 0 & 0 & 0 & 0 & 0 & 0 & 0 \\
                        0 & 0 & 0 & 0 & 0 & 0 & 0 & 0 & 0 \\
                        0 & 0 & 0 & 0 & 0 & 0 & 0 & 0 & 0 
                       \end{array} 
                \right) \delta (\omega - \tilde \omega),
\end{eqnarray}
which is a singular matrix. The zero modes of this matrix are $(\nu^{(0)})^T_1 = (0,0,\nu^{(0)}_{\bar P},0,0,0,0,0,0),$ 
$ (\nu^{(0)})^T_2 = (0,0,0,\nu^{(0)}_Q,0,0,0,0,0),$ $ (\nu^{(0)})^T_3 = (0,0,0,0,0,0,\nu^{(0)}_P,0,0), $
$ (\nu^{(0)})^T_4 = (0,0,0,0,0,0,0,\nu^{(0)}_{\bar Q},0)$ and $ (\nu^{(0)})^T_5 = (0,0,0,0,0,0,0,0,\nu^{(0)}_\rho)$, where
$ \nu^{(0)}_{\bar P}, \nu^{(0)}_P, \nu^{(0)}_Q, \nu^{(0)}_{\bar Q}, \nu^{(0)}_\rho$ are arbitrary constants. According to 
the Faddeev-Jackiw approach, these zero modes will lead to the following constraints of the theory, as 
\begin{eqnarray}
 \Omega^{(0)} & = & (\nu^{(0)})^T_A \; \frac{\partial V^{(0)}_J}{\partial \xi^A} \nonumber\\
 & = & \nu^{(0)}_{\bar P} \Big(\frac{P}{2} \Big) + \nu^{(0)}_{Q} \Big(\frac{\bar Q}{2 \phi^2} \Big) 
 + \nu^{(0)}_{P} \Big(\frac{1}{2} \bar P - \frac{1}{2 \phi} \phi_{\tilde z} \Big)
 + \nu^{(0)}_{\bar Q} \Big(\frac{1}{2 \phi^2} (Q - \rho_{\tilde z}) \Big) \approx 0.
\end{eqnarray}
In order to get further constraints we take the help of modified Faddeev-Jackiw approach, as described in the previous section. 
According to this approach, we have following combined equation;
\begin{eqnarray}
 f_{CD}^{(1)} \xi_z^D \; = \; Z_C (\xi), \label{fcdj}
\end{eqnarray}
where 
\begin{eqnarray}
  f_{CD}^{(1)} &=&  \left ( \begin{array}{c}
                          f_{AB}^{(0)} \\
                          \frac{\partial \Omega^{(0)}}{\partial \xi^A}
                         \end{array}
                 \right) 
              =  \left( \begin{array}{ccccccccc}
                        0 & -1 & 0 & 0 & 0 & 0 & 0 & 0 & 0 \\
                        1 & 0 & 0 & 0 & 0 & 0 & 0 & 0 & 0 \\
                        0 & 0 & 0 & 0 & 0 & 0 & 0 & 0 & 0 \\
                        0 & 0 & 0 & 0 & 0 & 0 & 0 & 0 & 0 \\
                        0 & 0 & 0 & 0 & 0 & -1 & 0 & 0 & 0 \\
                        0 & 0 & 0 & 0 & 1 & 0 & 0 & 0 & 0 \\
                        0 & 0 & 0 & 0 & 0 & 0 & 0 & 0 & 0 \\
                        0 & 0 & 0 & 0 & 0 & 0 & 0 & 0 & 0 \\
                        0 & 0 & 0 & 0 & 0 & 0 & 0 & 0 & 0 \\
                        0 & 0 & \frac{1}{2} & \frac{1}{2 \phi^2} & - \frac{\bar Q}{\phi^3} + \frac{\phi_{\tilde z}}{2 \phi^2}  - \frac{(Q - \rho_{\tilde z})}{\phi^3} & 0 & \frac{1}{2} & \frac{1}{2 \phi^2} & 0
                       \end{array} 
                \right) \delta (\omega - \tilde \omega), \nonumber\\
                {\text and} \nonumber\\
 Z_C (\xi) &=& \left(\begin{array}{c}
                      0 \\
                      0 \\
                      \frac{P}{2} \\
                      \frac{\bar Q}{2 \phi^2} \\
                      \frac{P \phi_{\tilde z}}{2 \phi^2} - \frac{(\phi_\omega \phi_{\tilde \omega} + \rho_{\tilde \omega} \bar \rho_\omega)}{\phi^3} - \frac{(Q \bar Q - \bar Q \rho_{\tilde z})}{\phi^3} \\
                      0 \\
                      \frac{\bar P}{2} - \frac{\phi_{\tilde z}}{2 \phi} \\
                      \frac{Q - \rho_{\tilde z}}{2 \phi^2} \\
                      0 \\
                      0
                     \end{array}
               \right).
\end{eqnarray}
The first-iterative symplectic matrix $f_{CD}^{(1)}$ is not a square matrix, however, it has following zero-modes;
$(\nu^{(1)})^T_1 = (0,0,0,0,0,0,0,0,\nu^{(1)}_\rho),$ $(\nu^{(1)})^T_2 = (0,0,-\frac{1}{\phi^2} \nu^{(1)}_{\bar P},0,0,0,0,\nu^{(1)}_{\bar Q},0),$ 
$(\nu^{(1)})^T_3 = (0,0,-\nu^{(1)}_{\bar P},0,0,0,0,\nu^{(1)}_{\bar Q},0),$ 
$(\nu^{(1)})^T_4 = (0,0,-\frac{1}{\phi^2} \nu^{(1)}_{\bar P},\nu^{(1)}_Q,0,0,0,0,0)$. Here $\nu^{(1)}_\rho,\nu^{(1)}_{\bar P}, \nu^{(1)}_{\bar Q},
\nu^{(1)}_Q $ are arbitrary constants. In order to deduce new constraints, we multiply these zeros modes to  
both sides of (\ref{fcdj}) and evaluate at $\Omega^{(0)} = 0$, i.e. 
\begin{eqnarray}
 Z_C (\xi) (\nu^{(1)})^T_C|_{\Omega^{(0)} = 0} = 0.
\end{eqnarray}
The above computation yields  $\Omega^{(1)} \equiv (\phi_\omega \phi_{\tilde \omega} + \rho_{\tilde \omega} \rho_\omega) \approx 0 $ as a new constraint. To check 
the existence of further constraints we repeat the same procedure as outlined above and calculate second-iterative symplectic matrix. With the 
help of the zero-modes of this second-iterative matrix it is easy to show that there are no further constraints in the system. 
The details of this computation is presented in the Appendix.

\section{Hamiltonian flow through hidden BRS invariance} 

 We want to discuss the Hamiltonian formulation for systems whose phase space is linear and are symplectic manifolds.
For that, let's begin with a symplectic vector space $(Z, \Omega)$. A vector field  $X:Z \to Z$ is called {\it Hamiltonian} if
\begin{eqnarray}
 \Omega^b(X(z)) = dH(z)
\label{om1}
\end{eqnarray}

$\forall z \in Z$  and for some $\mathbb{C}^1$ function $H : Z \to \mathbb{R}$. 

If such an $H$ exits we write the vector field as $X=X_H$ and call $H$ the Hamiltonian function for 
the vector field $X$. If $Z$ is finite-dimensional, as in our case, non-degeneracy of $\Omega$ implies that $\Omega^b : Z \to Z^*$ 
is an isomorphism, which further guarantees that $X_H$  exists for any given function $H$. However, if $Z$ is infinite-dimensional 
and $\Omega$ is only weakly non-degenerate, we do not know a priori whether $X_H$ exists or not for a given $H$. If it does, it is 
unique, since $\Omega^b$ is one-to-one.

The set of Hamiltonian vector fields on $Z$ is denoted by $\mathfrak{X}(Z)$. Thus $X_H \in \mathfrak{X}$ is a vector field determined 
by the following condition
\begin{eqnarray}
 \Omega(X_H(z), \delta z) = dH(z).\delta z \;\;\;\;\;\;   \forall z, \delta z \in Z
\label{om2}
\end{eqnarray}

If $X$ is a vector field the interior product $i_X\Omega$ is defined to be the dual vector (also called a one form) given at a point $z \in Z$ as :
$(i_X\Omega)_z \in Z^* $ ,  $(i_X\Omega)_z(v) := \Omega(X(z),v)$ $\forall v \in Z$, then condition (\ref{om1}) or (\ref{om2}) could be written as
\begin{eqnarray}
 i_X\Omega = dH
\label{bihamref}
\end{eqnarray}
Now one can use the vector field $X$ that generates the Hamiltonian flow to write the ASDYM field equations as 
\begin{eqnarray}
 X^A_z = {\bf X}_{K,J}(X^A),
\end{eqnarray}
where the subscripts $K$ and $J$ stand for Yang's $K$ and $J$ gauges respectively.
In the following we try to find an explicit expression for ${\bf X}_K$ and ${\bf X}_J$.

\subsection{K gauge}
In $K$ gauge the vector field ${\bf X}_K$ can be expressed simply as
\begin{eqnarray}
 {\bf X}_K = \tilde M_z \frac{\delta}{\delta \tilde M} + K_z \frac{\delta}{\delta K}.
\label{hamK}
\end{eqnarray}

The absence of the basis vector $\frac{\delta}{\delta M}$ in the expression of ${\bf X}_K$ can be ascribed to the vanishing of 
the corresponding coefficient which is again an artifact of the choice of a  local frame\footnote{ In this local frame, the ASD equations 
reduces to the simpler form as: 
$\partial_z \Phi_{\tilde z} - \partial_\omega \Phi_{\tilde \omega} = 0, 
\partial_{\tilde z} \Phi_{\tilde \omega} - \partial_{\tilde \omega} \Phi_{\tilde z} + [\Phi_{\tilde z}, \Phi_{\tilde\omega}] = 0.$} with  
$\Phi_\omega =  \Phi_z = 0$.  Another way to look at this is from the point of view of constraint analysis. From (\ref{cons}) it is clear
that $\Pi_{\tilde{M}}$ and $\Pi_K$ have non-vanishing canonical Poisson bracket between them and hence they are second-class constraints 
\`{a} la Dirac \cite{nut}, whereas $\Pi_M$ is a spurious first-class constraint and hence is prevented from playing any significant role 
in the phase space symplectic dynamics.

In early 90's Gozzi, Reuter and Thacker \cite{hidden1, hidden2, GRT} proposed a path integral formulation for classical Hamiltonian dynamics 
where they assumed a constant symplectic structure $\omega = \frac{1}{2} \omega_{AB} d\phi^A \wedge d\phi^B$ defined on symplectic phase space 
$\mathcal{M}_{2n}$, although in general $\omega$ is a closed 2 form ($d\omega = 0$) and non-degenerate ($det(\omega_{AB}) \neq 0$) everywhere 
in $\mathcal{M}_{2n}$. It has been shown how the classical path integral measure could be reformulated in terms of exponential of an action 
$\tilde{S}$ which not only depends on the bosonic phase space coordinates $\phi^A$ but also includes dual auxiliary 
variables $\lambda_A$ and fermionic anti-commuting ghosts $C^A$ and anti-ghosts $P_A$.

The ghosts $C^A$ are interpreted  as one forms $d\phi^A$ \cite{GRT} that construct the basis of cotangent space 
$T^*_\phi \mathcal{M}_{2n}$, whereas the anti-ghosts $P_A$ constitute the basis of tangent space $T_\phi \mathcal{M}_{2n}$. With this 
new formalism one can have a unified framework for evolution of scalars as well as p-form densities on phase space. The classical path 
integral variables $(\phi, \lambda)$ and $(C, P)$ form canonically conjugate pairs and the only non-vanishing equal-time graded commutation
relations satisfied by them are 

\begin{eqnarray}
\big[\hat{\phi}^A , \hat{\lambda}_B\big] = i \delta^A_B \nonumber\\
\big[\hat{C}^A, \hat{P}_B\big] = \delta^A_B. 
\label{com}
\end{eqnarray}

The commutators defined in (\ref{com}) clearly hints at interpreting variables $\phi^A$, $\lambda_A$, $C^A$, $P_A$ as coordinates of 
the $4 \times (2n)$ - dimensional extended phase space on which the graded Poisson structure are defined. One can derive the 
graded Poisson structure by considering $\lambda_A$ and $P_A$ as the constraints derived from the first-order Lagrangian which are namely
$\Pi_{\phi^A} = \lambda_A$ and $\Pi_{C^A} = iP_A$ where $\Pi_{\phi^A}$ and $\Pi_{C^A}$ are the momenta conjugate to $\phi^A$ and $C^A$. 
Due to the presence of these constraints, the passage from Lagrangian to Hamiltonian formalism calls for application of Dirac procedure 
\cite{dirac} where the Dirac bracket is given by

\begin{eqnarray}
\big\{\hat{\phi}^A , \hat{\lambda}_B\big\}_D  = \delta^A_B \nonumber\\
\big\{{\hat{C}^A, \hat{P}_B}\big\}_D = i\delta^A_B. 
\label{dirac}
\end{eqnarray}

From (\ref{com}) it is clear that the extended phase space variables in $K$ gauge can be directly written in the Schr\"{o}dinger representation 
from the corresponding hatted operator as follows. The Grassmann parity zero variables are 

\begin{eqnarray}
&& \hat{\phi}^A = \phi^A, \;\;\; \forall A =1,2 \nonumber\\
&& \hat{\phi}^1 = \phi^1 = \tilde M, \qquad
\hat{\phi}^2 = \phi^2 = K,
\end{eqnarray}

and \begin{eqnarray}
&& \hat{\lambda}_A = - \frac{\partial}{\partial \phi^A}, \nonumber\\
&& \hat{\lambda}_1 = - \frac{\partial}{\partial \tilde M}, \qquad
\hat{\lambda}_2 = - \frac{\partial}{\partial K},  
\end{eqnarray}

whereas the Grassmann parity one variables are 

\begin{eqnarray}
&& \hat{ C}^A = C^A = d \phi^A, \nonumber\\
&& \hat{ C}^1 = C^1 = d \phi^1 = d \tilde M, \qquad
\hat{ C}^2 = C^2 = d \phi^2 = d K, 
\end{eqnarray}

and

\begin{eqnarray}
&& \hat{P}_A = \frac{\partial}{\partial C^A} = \frac{\partial}{\partial \phi^A}, \nonumber\\
&& \hat{P}_1 = \frac{\partial}{\partial C^1} = \frac{\partial}{\partial \phi^1} = \frac{\partial}{\partial \tilde M}, \qquad
\hat{P}_2 = \frac{\partial}{\partial C^2} = \frac{\partial}{\partial \phi^2} = \frac{\partial}{\partial K}.
\end{eqnarray}

The action $\tilde{S}$ is invariant under a set of transformations generated by the conserved charges \cite{hidden2} namely the symplectic 
2-form $\omega = \frac{1}{2} \omega_{AB}C^A C^B$, the symplectic bi-vector $\Omega = \frac{1}{2} \omega^{AB}P_A P_B$ and their conservation 
is nothing but the Liouville theorem in classical mechanics.

Operationally the BRS (Becchi-Rouet-Stora) like charge $(\hat{Q})$ which acts as the exterior derivative on phase space can be defined as 
\begin{eqnarray}
 \hat{Q} &=& \hat{C}^A \hat{\lambda}_A \nonumber\\
 &=& \hat{C}^1 \hat{\lambda}_1 +  \hat{C}^2 \hat{\lambda}_2  \nonumber\\
 &=& - \Big(d \tilde M \frac{\partial}{\partial \tilde M} + d K \frac{\partial}{\partial K}\Big),
\end{eqnarray}
 while the anti-BRS like charge $(\hat{\bar Q}_r)$  $\forall r = 0,1$ which is like exterior co-derivative mapping p-vectors to 
 (p+1)-vectors,  can be given as 
 \begin{eqnarray}
 \hat{\bar Q}_r &=& \hat{P}_A \; \omega_r^{AB} \; \hat{\lambda}_B  \nonumber\\
 \hat{\bar Q}_{0,1} &=& \hat{P}_1 \; \omega_{0,1}^{12} \; \hat{\lambda}_2 + \hat{P}_2 \; \omega_{0,1}^{21} \; \hat{\lambda}_1. 
\end{eqnarray}

And then there are ghost charges $Q_{ghost} = C^A P_A$ which counts the form or vector number attaching a weight +1 to each one form $C^A$ 
and -1 to each tangent vector $P_A$. The BRS and anti-BRS like charges are nilpotent and anti-commute with each other.

\begin{eqnarray}
\big\{ \hat{Q},  \hat{Q}\big \} = 0, \;\;  \big\{ \hat{\bar Q}_r ,  \hat{\bar Q}_r \big \} = 0, \;\;\; \big \{ \hat{Q}, \hat{\bar Q}_r\big \} = 0.
\end{eqnarray}

For more details regarding group theory within the BRS-anti BRS framework, $Q$-bracket on supersymplectic manifold, readers are directed 
to \cite{mar}.

It must be noted while we are using index r in the definition of anti-BRS charges we are implicitly assuming the existence of two constant 
symplectic 2-forms that the symplectic phase space $\mathcal{M}_{2n}$ is endowed with, which in turn guarantees the existence of bi-Hamiltonian 
structure i.e. one can associate two different Hamiltonian functions wrt to these two symplectic structures but the pairs $(H_1(\phi), \omega_1)$ 
and $(H_2(\phi), \omega_0)$ give rise to the identical equations of motion which is 

\begin{eqnarray}
\dot{\phi}^A = \omega^{AB}_1 \partial_B H_1 = \omega^{AB}_0 \partial_B H_2.
\label{biHK}
\end{eqnarray}
 
Geometrically the equation (\ref{biHK}) can be interpreted as generating a Hamiltonian vector field $X$ and the corresponding 
flow equation can have two equivalent form (or Hamiltonian descriptions) wrt anti BRS like charges namely
\begin{eqnarray}
 {\bf X_K} = [\hat{H}_2, \hat{\bar Q}_0] = [\hat{H}_1, \hat{\bar Q}_1],
\end{eqnarray}
since in component form \cite{silka} using (\ref{EOMK}) one can write down ${\bf X_K}$ as
\begin{eqnarray}
{\bf X_K} &=& [\hat{H}_1, \hat{\bar Q}_1] = \hat{P}_A \; \omega_1^{AB} \partial_B \hat{H}_1 = \hat{P}_A \dot{\hat{\phi}}^A \nonumber\\
 &=& \hat{P}_1 {\hat{\phi}}^1_z + \hat{P}_2 {\hat{\phi}}^2_z \nonumber\\
 &=& \frac{\partial}{\partial \tilde M} \tilde M_z + \frac{\partial}{\partial K} K_z \nonumber\\
&=& \left(M_{\tilde z} + \frac {2}{3} ([M_\omega, K] + [K_\omega, M])\right) \frac{\partial}{\partial \tilde M} + M \frac{\partial}{\partial K}.
\label{HamK}
\end{eqnarray}

One can see the end result of equations (\ref{hamK}) and (\ref{HamK}) are identical by construction.

\subsection{J gauge}
Now we can repeat the same strategy as in {\it K} gauge to explore the Hamiltonian flow in phase space through hidden BRS invariance
\cite{GRT}

\vspace{0.5cm}

In J gauge, there are four extended phase space bosonic variables which are

\begin{eqnarray}
&& \hat{\phi}^A = \phi^A, \;\;\; \forall A = 1,2,3,4 \nonumber\\
&& \hat{\phi}^1 =  \phi^1 = \bar \rho, \quad \hat{\phi}^2 =  \phi^2 = \bar P, \quad 
\hat{\phi}^3 =  \phi^3 = Q, \quad \hat{\phi}^4 = \phi^4 = \phi,
\end{eqnarray}

and then there are four corresponding dual auxiliary variables which are

\begin{eqnarray}
&& \hat{\lambda}_A = - \frac{\partial}{\partial \phi^A}, \nonumber\\
&& \hat{\lambda}_1 = - \frac{\partial}{\partial \bar \rho}, \quad  
\hat{\lambda}_2 = - \frac{\partial}{\partial \bar P}, \quad 
\hat{\lambda}_3 = - \frac{\partial}{\partial Q}, \quad 
\hat{\lambda}_4 = - \frac{\partial}{\partial \phi},  
\end{eqnarray}

whereas the fermionic variables are four anticommuting ghosts

\begin{eqnarray}
&& \hat{ C}^A = C^A = d \phi^A, \nonumber\\
&& \hat{ C}^1 = d \bar \rho, \quad \hat{ C}^2 = d \bar P, \quad 
\hat{ C}^3 = d Q, \quad \hat{ C}^4 = d \phi,
\end{eqnarray}

and four anti-ghosts which are

\begin{eqnarray}
&& \hat{P}_A = \frac{\partial}{\partial C^A} = \frac{\partial}{\partial \phi^A} =  - \hat{\lambda}_A,  \nonumber\\
&& \hat{P}_1 =  \frac{\partial}{\partial \bar \rho}, \quad  
\hat{P}_2 =  \frac{\partial}{\partial \bar P}, \quad 
\hat{P}_3 =  \frac{\partial}{\partial Q}, \quad 
\hat{P}_4 =  \frac{\partial}{\partial \phi}.  
\end{eqnarray}

The natural question that crops up in our mind at this stage is why the variables $P$, $\bar Q$ and $\rho$ do not take part in the 
symplectic dynamics and the answer is straightforward and can be verified from (\ref{consj}). It is instructive to note that 
the pairs  ($\Pi_{\bar \rho}$ , $\Pi_Q$) and  ($\Pi_{\phi}$ , $\Pi_{\bar {P}}$) have non-vanishing Poisson brackets among them
and hence can be classified as second-class constraints \cite{nut} whereas following three constraints namely 
$\Pi_P$, $\Pi_{\bar{Q}}$ and $\Pi_{\rho}$ are spurious first-class constraints which we will ignore and hence the 
corresponding  variables $P$, $\bar{Q}$ and $\rho$ do not play any role in the Hamiltonian flow.

Similar to $K$ gauge, the BRS (Becchi-Rouet-Stora) like charge $(\hat{Q})$ which acts as the exterior derivative on phase space 
can be defined in $J$ gauge  as well
\begin{eqnarray}
 \hat{Q} &=& \hat{C}^A \hat{\lambda}_A \nonumber\\
 &=& \hat{C}^1 \hat{\lambda}_1 +  \hat{C}^2 \hat{\lambda}_2  + \hat{C}^3 \hat{\lambda}_3 + \hat{C}^4 \hat{\lambda}_4 \nonumber\\
 &=& - \Big(d \bar \rho \frac{\partial}{\partial \bar \rho} + d \bar P \frac{\partial}{\partial \bar P} +  d Q \frac{\partial}{\partial Q} 
 +  d \phi \frac{\partial}{\partial \phi}\Big),
\end{eqnarray}
 while the anti-BRS like charge $(\hat{\bar Q}_r)$ which is like exterior co-derivative is given by 
 \begin{eqnarray}
 \hat{\bar Q}_r &=& \hat{P}_A \; \omega_r^{AB} \; \hat{\lambda}_B  \nonumber\\
 \hat{\bar Q}_{0,1} &=& \hat{P}_1 \; \omega_{0,1}^{12} \; \hat{\lambda}_2 + \hat{P}_2 \; \omega_{0,1}^{21} \; \hat{\lambda}_1. 
\end{eqnarray}

The integrability and thus the  bi-Hamiltonian structure of ASDYM in $J$ gauge enables us to write two different 
Hamiltonian functions wrt two symplectic structures with a condition that the pairs $(H_1(\phi), \omega_1)$ and $(H_2(\phi), \omega_0)$ 
give same equations of motion

\begin{eqnarray}
\dot{\phi}^A = \omega^{AB}_1 \partial_B H_1 = \omega^{AB}_0 \partial_B H_2.
\label{biHJ}
\end{eqnarray}

This generates a Hamiltonian vector field $X$ and the corresponding flow equation can be described in two equivalent ways wrt anti BRS 
like charges in $J$ gauge namely
\begin{eqnarray}
 {\bf X_J} = [\hat{H}_2, \hat{\bar Q}_0] = [\hat{H}_1, \hat{\bar Q}_1],
\end{eqnarray}
since in component form \cite{silka} using (\ref{EOMJ}) one can write down ${\bf X_J}$ as
\begin{eqnarray}
 {\bf X_J} &=& [\hat{H}_1, \hat{\bar Q}_1] = \hat{P}_A \; \omega_1^{AB} \partial_B \hat{H}_1 = \hat{P}_A \dot{\hat{C}}^A \nonumber\\
 &=& \hat{P}_1 {\hat{\phi}}^1_z + \hat{P}_2 {\hat{\phi}}^2_z +  \hat{P}_3 {\hat{\phi}}^3_z +  \hat{P}_4 {\hat{\phi}}^4_z\nonumber\\
 &=& \frac{\partial}{\partial \bar \rho} \bar \rho_z + \frac{\partial}{\partial \bar P} \bar P_z  + \frac{\partial}{\partial Q} Q_z + \frac{\partial}{\partial \phi} \phi_z \nonumber\\
 &=&  \bar Q \frac{\partial}{\partial \bar \rho} + \left( -P_{\tilde z} + 2 \phi^{-1} \phi_{\omega \tilde\omega} - 2 \phi^{-2} \phi_\omega \phi_{\tilde \omega} + 2 \phi^{-2} \rho_{\tilde \omega} \bar \rho_\omega
- 2 \phi^{-2} Q\bar Q \right) \frac{\partial}{\partial \bar P}  \nonumber\\
&+& \left( \phi^2(\phi^{-2} \rho_{\tilde \omega})_\omega - 2PQ \right) \frac{\partial}{\partial Q} + P \phi \frac{\partial}{\partial \phi},  
\label{HamJ}
\end{eqnarray}
which is identical to what has been found in \cite{nut}.

\subsection{Bi-Hamiltonian Structure and Compatibility}

In this subsection, we make some remarks on the bi-Hamiltonian nature of ASDYM system in Yang's $J$ and $K$ gauges as was already observed in previous two subsections. 
Our symplectic manifold $\mathcal{M}_{2n}$ is endowed with a Hamiltonian vector field ${\bf X}$ with respect to two different symplectic 
structures~$ \omega _1 $ and $~\omega _0$, that is (see eq. (\ref{bihamref})),
\begin{eqnarray}
    \mathbf{i} _{ X } \omega _1 = dH_1 \;\;\; \text{and} \;\;\; \mathbf{i} _{ X } \omega _0 = dH_2,
\label{bihammaster}
\end{eqnarray}
where $ H_1 $ and $ H_2 $ are two distinct, Hamiltonian functions. At this point, we introduce the so called
{\it recursion operator}  $ Z = \omega _0 ^\sharp \omega _1 ^\flat \colon TM \to TM $, where  $ \omega ^\flat \colon TM \to T ^\ast M $ denotes 
the ``musical'' isomorphism induced by the symplectic form $ \omega $, and $ \omega ^\sharp $ is its inverse.  Consequently, $ X $ being 
Hamiltonian vector field with respect to two symplectic forms, the associated Hamiltonian flow preserves the eigenvalues of the 
recursion operator $Z$. Hence, if $ Z $ has  $n $  functionally independent eigenvalues which are in involution, then, it 
guarantees complete integrability according to Liouville--Arnold theorem.

In order to find integrability in this situation one is naturally compelled to try to find out sufficient conditions 
for eigenvalues of $Z$ to be in involution. Several conditions of this type do already exist in literature.

One such condition is  due to the pioneering work by Magri \cite{Magri_1978} in infinite-dimensional case.  It was proved by 
Magri and Morosi \cite{Magri_1984} that, if  the sum of the Poisson tensors associated respectively to $ \omega _1 $ and $ \omega _0 $ is 
still a Poisson tensor, then the eigenvalues of $Z$ are in involution. In this case we say that $ \omega _1 $ and $ \omega _0 $ are 
{\it Magri-compatible} and  the triple $ (\mathcal{M}, \omega _1 , \omega _0) $ is a {\it bi-Hamiltonian manifold}, and the quadruple  
$ (\mathcal{M} , \omega _1 , \omega _0 , X) $  is a {\it bi-Hamiltonian system} \`{a} la Magri if there exist functions 
$H _1$ and $H _2$ such that $ X = \omega _1 ^\sharp  \cdot  dH _1 = \omega _0 ^\sharp \cdot   dH _2$. 

It must be noted that not all completely integrable Hamiltonian systems are bi-Hamiltonian \`{a} la Magri.  Brouzet \cite{Brouzet_1990} and 
Fernandes \cite{Fernandes_1994} found that there exist completely integrable systems that are not bi-Hamiltonian.  
This limitation in Magri's approach led others explore different notions of compatibility
as far as integrability is concerned.

In the literature of integrable systems, there is a sufficient condition dubbed as {\it strong dynamical compatibility} \`{a} la 
Bogoyavlenskij \cite{Bogoyavlenskij_1996}. This condition demands the existence of a Hamiltonian vector field $X$ wrt two 
symplectic structures~$ \omega _1 $ and~$ \omega _0 $ which is completely integrable wrt $ \omega _1 $, and  is 
{\it non-degenerate} i.e. the orbits of $X$ lie on the Lagrangian tori and  in any local $ \omega _1 $-action-angle 
coordinates $ (a, \alpha)$, the Hamiltonian $ H _1 $ of $X$  associated to $ \omega _1 $ satisfies the following Hessian condition
\begin{eqnarray}
    \det \left( \frac{ \partial ^2 H _1 } { \partial a _i \partial a _j } \right)(a) \neq 0.
\end{eqnarray}

There is yet another notion of compatibility, that was first introduced by Fass\`o and Ratiu \cite{Fasso_1998} in order to 
study superintegrable systems  with motions  constrained on isotropic tori of dimension less than $ n $, instead of 
Lagrangian tori of dimension $n$. If we assume $\omega _1$ and  $\omega _0$ to be two symplectic forms on $\mathcal{M}$, the 
fibration (foliation) is said to be bi-Lagrangian if the fibers (leaves) are Lagrangian wrt both  $\omega _1$ and  $\omega _0$. 
And if there exist a bi-Lagrangian fibration of $ \mathcal{M} $, we infer that $ \omega _1 $ and $ \omega _0 $ are {\it bi-affinely compatible}  
if the Bott connection associated to $\omega _1$ and $\omega _0$ coincide with each other.

\section{Conclusions}

In our present endeavor, we have derived the full set of constraints for a system of ASDYM equations recast into a Lagrangian formalism, 
in $K$ as well as $J$ gauges, by means of the modified Faddeev-Jackiw method which combines the usual 
Faddeev-Jackiw approach with the Dirac-Bergmann algorithm. In the modified method, to derive new constraints, a consistency condition 
analogous to the Dirac-Bergmann algorithm is used which makes this method economical and  at the same time convenient to deal with  systems 
having constraints. 

We have aptly used the hidden BRS invariance and extended phase space formalism to derive the Hamiltonian flow for ASDYM system in both J and K gauges. 
Our results match with previous findings \cite{nut} where Dirac constraint analysis and symplectic techniques were used to constrain the phase 
space variables and write down the Hamiltonian flow.

Finally let us briefly mention some possible further directions.
Yang found that SU(2) SDYM equations in a particular gauge choice are related to a principal chiral model on a four dimensional flat 
submanifold $X$ of $\mathbb{R}^4$ \cite{yangPRL, plebanskiPLA}. Interestingly for K\"{a}hler and hyper-K\"{a}hler manifolds, {\it Yang equations} admit 
a natural extension to the so called {\it Donaldson-Nair-Schiff equation} \cite{donald, NS} which represents a coupling between ASDYM and anti 
self-dual gravity (ASDG). It would be instructive to perform a full constraint analysis 
\`{a} la modified Faddeev-Jackiw approach for ASDG with topological terms \cite{monte}. 
The constraint analysis can further test the appropriateness of the analogy between $\theta$-angle in Yang-Mills and 
the Barbero-Immirzi parameter in gravity. Another worthwhile exercise could be 
to start with the new BF-type first order Lagrangian found in \cite{kras} interpolating between topological and 
anti-self dual gravity and perform constraint analysis. Some of these issues are under investigation and will be reported elsewhere.

\section*{Acknowledgements}
The research work of S.G. is supported by the Conselho Nacional de Desenvolvimento Cient\'{i}fico e Tecnol\'{o}gico (CNPq), 
Brazil, Grant No. 151112/2014-2. The research work of R.R. was supported by FAPESP through Instituto de Fisica, Universidade de Sao Paulo 
with grant number 2013/17765-0.

\section*{Appendix}
In this appendix, we provide explicit calculation for the second-iterative symplectic matrix and derivation of new constraints in $J$-gauge. 
The second-iterative symplectic matrix $f_{EF}^{(2)}$ can be given as follows
\begin{eqnarray}
 f_{EF}^{(2)} &=&  \left ( \begin{array}{c}
                          f_{CD}^{(1)} \\
                          \frac{\partial \Omega^{(1)}}{\partial \xi^A}
                         \end{array}
                 \right) 
              =  \left( \begin{array}{ccccccccc}
                        0 & -1 & 0 & 0 & 0 & 0 & 0 & 0 & 0 \\
                        1 & 0 & 0 & 0 & 0 & 0 & 0 & 0 & 0 \\
                        0 & 0 & 0 & 0 & 0 & 0 & 0 & 0 & 0 \\
                        0 & 0 & 0 & 0 & 0 & 0 & 0 & 0 & 0 \\
                        0 & 0 & 0 & 0 & 0 & -1 & 0 & 0 & 0 \\
                        0 & 0 & 0 & 0 & 1 & 0 & 0 & 0 & 0 \\
                        0 & 0 & 0 & 0 & 0 & 0 & 0 & 0 & 0 \\
                        0 & 0 & 0 & 0 & 0 & 0 & 0 & 0 & 0 \\
                        0 & 0 & 0 & 0 & 0 & 0 & 0 & 0 & 0 \\
                        0 & 0 & \frac{1}{2} & \frac{1}{2 \phi^2} & - \frac{\bar Q}{\phi^3} + \frac{\phi_{\tilde z}}{2 \phi^2}  - \frac{(Q - \rho_{\tilde z})}{\phi^3} & 0 & \frac{1}{2} & \frac{1}{2 \phi^2} & 0 \\
                        0 & 0 & 0 & 0 & \frac{-3}{\phi^4} (\phi_\omega \phi_{\tilde \omega} + \rho_{\tilde \omega} \bar \rho_\omega) & 0 & 0 & 0 & 0
                       \end{array} 
                \right) \delta (\omega - \tilde \omega). \nonumber\\
\end{eqnarray}
As one can easily see  the above second-iterative matrix is not a square matrix yet it has following zero modes: 
$(\nu^{(2)})^T_1 = (0,0,0,0,0,0,0,0,\nu^{(2)}_\rho),$ 
$(\nu^{(2)})^T_2 = (0,0,-\frac{1}{\phi^2} \nu^{(2)}_{\bar P},0,0,0,0,\nu^{(2)}_{\bar Q},0),$ 
$(\nu^{(2)})^T_3 = (0,0,-\nu^{(2)}_{\bar P},0,0,0,0,\nu^{(2)}_{\bar Q},0),$ 
$(\nu^{(2)})^T_4 = (0,0,-\frac{1}{\phi^2} \nu^{(2)}_{\bar P},\nu^{(2)}_Q,0,0,0,0,0)$, where 
$\nu^{(2)}_\rho, \nu^{(2)}_{\bar P}, \nu^{(2)}_{\bar Q}, \nu^{(2)}_Q $ are arbitrary constants. According to the modified 
Faddeev-Jackiw approach the second-iterative combined equation reads as 
\begin{eqnarray}
 f_{EF}^{(2)} \; \xi_z^F \; = \; Z_E (\xi),
\end{eqnarray}
with 
\begin{eqnarray}
 Z_E (\xi) &=& \left(\begin{array}{c}
                      0 \\
                      0 \\
                      0 \\
                      0 \\
                      - \frac{(\phi_\omega \phi_{\tilde \omega} + \rho_{\tilde \omega} \bar \rho_\omega)}{\phi^3} \\
                      0 \\
                      0 \\
                      0 \\
                      0 \\
                      0 \\
                      0
                     \end{array}
               \right).
\end{eqnarray}
In order to get new constraints, if any, we calculate 
\begin{eqnarray}
Z_E (\xi) (\nu^{(2)})^T_E|_{\Omega^{(1)} = 0} = 0,
\end{eqnarray}
which yields identities, i.e. $0 = 0$, on the constraint surface $\Omega^{(1)} = 0$. Thus, we have no further constraints in the system.

\end{document}